\documentclass[a4paper,fleqn,usenatbib]{mnras}
\usepackage{amsmath}
\usepackage{mathptmx}
\usepackage{txfonts}
\usepackage[T1]{fontenc}
\usepackage{ae,aecompl}
\usepackage{graphicx}	
\usepackage{amssymb}	

\newcommand{\e}{\end{equation}}
\newcommand{\bear}{\begin{eqnarray}}
\newcommand{\ear}{\end{eqnarray}}
\def\aj{AJ}
\def\apj{ApJ}
\def\apjs{ApJS}
\def\jcap{JCAP}
\def\mnras{MNRAS}
\def\aap{A\&A}
\def\prd{Physical Review D}
\def\nat{Nature}
\def\apjs{ApJS}
\def\apjl{ApJ Letters}
\def\physrep {Physics Reports}

\title[Testing homogeneity in SDSS DR12 with Shannon entropy]{Testing
  homogeneity in the Sloan Digital Sky Survey Data Release Twelve with
  Shannon entropy}
    
\author[Pandey, B. and Sarkar, S.]  {Biswajit
  Pandey\thanks{E-mail: biswap@visva-bharati.ac.in} and Suman
  Sarkar\thanks{E-mail:suman2reach@gmail.com} \\
  Department of Physics, Visva-Bharati University, Santiniketan,
  Birbhum, 731235, India\\ }
       
 \date{\today}

 \pubyear{2015}
  
\begin{document}
\label{firstpage}
\pagerange{\pageref{firstpage}--\pageref{lastpage}}      
\maketitle
       
 \begin{abstract}
       
 We analyze a set of volume limited samples from SDSS DR12 to quantify
 the degree of inhomogeneity at different length scales using Shannon
 entropy. We find that the galaxy distributions exhibit a higher
 degree of inhomogeneity as compared to a Poisson point process at all
 length scales. Our analysis indicates that signatures of
 inhomogeneities in the galaxy distributions persist at least upto a
 length scale of $120 \, h^{-1}\, {\rm Mpc}$. The galaxy distributions
 appear to be homogeneous on a scale of $140 \, h^{-1}\, {\rm Mpc}$
 and beyond. Analyzing a set of mock galaxy samples from a semi
 analytic galaxy catalogue from the Millennium simulation we find
 a scale of transition to homogeneity at $\sim 100 \, h^{-1}\,
 {\rm Mpc}$.
       \end{abstract}

       
       \section{Introduction}

Homogeneity and isotropy on sufficiently large scales is an assumption
which is fundamental to our current understanding of the
Universe. This assumption which is popularly known as the Cosmological
principle implies that the properties of the Universe is same for all
observers irrespective of their locations and the directions at which
they are looking at. The assumption has also an aesthetic appeal of
its own as it indirectly asserts us that the same physical laws must
apply throughout the Universe given it is homogeneous and isotropic to
start with. One can not prove the Cosmological principle in a strictly
mathematical sense but fortunately one can test it against the
presently available cosmological observations. The existence of the
cosmic microwave background radiation and its near uniform temperature
over the entire sky provides by far the best conclusive evidence in
favour of isotropy \citep{penzias,smoot,fixsen}. Other important
supporting evidences come from the isotropy in angular distributions
of radio sources \citep{wilson,blake}, isotropy in the X-ray
background \citep{peeb93,wu,scharf} and the isotropy in the
distribution of neutral hydrogen as shown by lyman-$\alpha$
transmitted flux \citep{hazra}. But mere existence of isotropy around
us does not automatically guarantees homogeneity of the
Universe. Spatial homogeneity can be asserted from isotropy only when
the later is confirmed around each point with the hypothesis that the
matter distribution is a smooth function of position \citep{straumann,
  labini10}.

We see structures in the present Universe starting from planets, stars
and galaxies to groups, clusters and superclusters spanning a wide
range of length scales. The present Universe looks highly
inhomogeneous on small scales and it is important to test if such
inhomogeneities continue to persist on large scales. Large scale
inhomogeneities have several important implications for cosmology and
could pose a serious challenge to the standard cosmological
framework. Inhomogeneities, through the backreaction mechanism can
provide an alternate explanation of a global cosmic acceleration
without requiring any additional dark energy component
\citep{buchert97, buchert01, schwarz, kolb06, paranjape09, kolb10,
  ellis}. A large void can also mimic an apparent acceleration of
expansion \citep{tomita01}. Fortunately such models can be constrained
with observations such as SNe, CMB, BAO and measurements of angular
diameter distance and Hubble parameter \citep{zibin, clifton, biswas,
  chris, larena, valken}. However if our Universe is inhomogeneous on
large scales, the currently available inhomogeneous cosmological
models suggest that the effects of inhomogeneities on observed
quantities are non-negligible and more precise observations will not
be properly analyzed unless inhomogeneities are taken into account
\citep{bolejko}.

Galaxy surveys map the three dimensional distribution of galaxies in
the Universe. If our Universe is homogeneous then the statistical
properties of galaxy distributions in a finite volume should be
independent of the location of that volume in the Universe. One simple
tool to characterize the statistical properties of galaxy
distributions is the two point correlation function \citep{peeb80}
which measures the excess probability of finding a pair of points
separated by a distance $r$ compared with a random Poisson
process. The two point correlation function on small scales $ 0.1 \,
h^{-1} \rm {Mpc} \leq r \leq 10 \, h^{-1} \rm {Mpc}$, is well
described by a power law of the form $\xi(r)
=(\frac{r}{r_{0}})^{-\gamma}$, with correlation length $r_{0} \sim
5-6\,h^{-1} \rm {Mpc}$ and slope $\gamma \sim 1.7-1.8$
\citep{hawkins,zehavi05}.  $\xi(r)$ vanishes at scales $> 20 \, h^{-1}
\rm {Mpc}$ which is consistent with large scale homogeneity.  However
the problem with correlation function analysis is that it assumes a
mean density on the scale of survey which is not a defined quantity
below the scale of homogeneity. Most of the statistical tests of
homogeneity carried out so far are based on the count in spheres
method where the number counts $n(<r)$ within a sphere of radius $r$
and its scaling with $r$ is used to test the transition scale to
homogeneity. If the scaling exponent approaches a value $\sim 3$ on
some scale it marks the transition to homogeneity indicating that the
distribution is homogeneous on and above that scale \citep{ hogg,
  labini11a, scrim}. Fractal analysis \citep{martinez90, coleman92,
  borgani95, bharad99, yadav10} uses the scaling of different moments
of $n(<r)$ to characterize the scale of homogeneity. Various studies
carried out with these methods claim to have found a transition to
homogeneity on sufficiently large scales $70-150 \, h^{-1} \rm {Mpc}$
\citep{martinez94, guzzo97, martinez98, bharad99, pan2000, kurokawa,
  hogg, yadav, prksh2, scrim} whereas there are studies which claim
the absence of any such transition out to the scale of the survey
\citep{coleman92, amen, labini07, labini09a, labini09b,
  labini11b}. The present generation galaxy surveys (SDSS,
\citealt{york}; 2dFGRS, \citealt{colles}) have now mapped the
distribution of millions of galaxies across billions of light years
providing an unique opportunity to test the assumption of homogeneity
on large scales. The Sloan Digital Sky Survey (SDSS) has reached its
final stage (Data Release 12) \citep{alam} now encompassing more than
one-third of the entire celestial sphere. This provides us with galaxy
distributions in enormous volumes of the Universe for which the
statistical properties of galaxy distributions can be analyzed and
homogeneity of the Universe can be tested with greater confidence.
\citet{pandey} introduce a method based on Shannon entropy
\citep{shannon48} for characterizing inhomogeneities and applied the
method on some Monte Carlo simulations of inhomogeneous distributions
and N-body simulations which show that the proposed method has great
potential for testing the large scale homogeneity in galaxy redshift
surveys.

In the present work we employ the method proposed by \citet{pandey} to
analyze a set of volume limited samples from the Sloan Digital Sky
Survey Data Release Twelve (SDSS DR12). We quantify the
inhomogeneities in the galaxy distribution using Shannon entropy and
test if there is a scale of transition to homogeneity. We also analyze
a set of mock galaxy samples from a semi analytic galaxy catalogue
\citep{guo} derived from the Millennium Run simulation
\citep{springel} to compare theoretical predictions with observations.

A brief outline of the paper follows. We briefly describe our method
in Section 2, describe the Monte Carlo simulation, SDSS and Millennium
data in Section 3 and present the results and conclusions in Section
4.

We have used a $\Lambda$CDM cosmological model with $\Omega_{m0}=0.3$,
$\Omega_{\Lambda0}=0.7$ and $h=1$ throughout.
       
\section{METHOD OF ANALYSIS}

\citet{pandey} propose a method based on the Shannon entropy to study
inhomogeneities in a 3D distribution of points. Shannon entropy
\citep{shannon48} is originally proposed by Claude Shannon to quantify
the information content in strings of text. It gives a measure of the
amount of information required to describe a random variable. The
Shannon entropy for a discrete random variable $X$ with $n$ outcomes
$\{x_{i}:i=1,....n\}$ is a measure of uncertainty denoted by $H(X)$
defined as,

\begin{equation}
H(X) =  - \sum^{n}_{i=1} \, p(x_i) \, \log \, p(x_i)
\label{eq:shannon1}
\end{equation}

where $p(x)$ is the probability distribution of the random variable
$X$.  

 Given a set of $N$ points distributed in 3D we consider each of the
 $i^{th}$ points as center and determine $n_i(<r)$ the number of other
 points within a sphere of radius $r$ as,\\
\begin{equation}
 n_i(<r)=\sum_{j=1}^{N}\Theta(r-\mid {\bf{x}}_i-{\bf{x}}_j \mid)
\label{eq:count}
\end{equation}
where $\Theta$ is the Heaviside step function and ${\bf{x}_{i}}$ and
${\bf{x}_{j}}$ are the radius vector of $i^{th}$ and $j^{th}$ points
respectively. To avoid any edge effects we discard all the centers
which lie within a distance $r$ from the survey boundary. The number
of available centers will decrease with increasing $r$ for any finite
volume sample. We define a separate random variable $X_{r}$ for each
radius $r$ which has $M(r)$ possible outcomes each given by,
$f_{i,r}=\frac{n_{i}(<r)}{\sum^{M(r)}_{i=1} \, n_{i}(<r)}$ with the
constraint $\sum^{M(r)}_{i=1} \, f_{i,r}=1$. 
The Shannon entropy associated with the random variable $X_{r}$ can be
written as,
\begin{eqnarray}
H_{r}& = &- \sum^{M(r)}_{i=1} \, f_{i,r}\, \log\, f_{i,r} \nonumber\\ &=& 
\log(\sum^{M(r)}_{i=1}n_i(<r)) - \frac {\sum^{M(r)}_{i=1} \,
  n_i(<r) \, \log(n_i(<r))}{\sum^{M(r)}_{i=1} \, n_i(<r)}
\label{eq:shannon2}
\end{eqnarray}
Where the base of the logarithm is arbitrary and we choose it to be
$10$. $f_{i,r}$ will have the same value $\frac{1}{M(r)}$ for all the
centers when $n_{i}(<r)$ is same for all of them. This is an ideal
situation when the spheres of radius $r$ around each of the $M(r)$
centres contain exactly the same number of points. This maximizes the
Shannon entropy to $(H_{r})_{max}=\log \, M(r)$ for radius $r$. We
define the relative Shannon entropy as the ratio of the entropy of a
random variable $X_{r}$ to the maximum possible entropy
$(H_{r})_{max}$ associated with it. The relative Shannon entropy
$\frac{H_{r}}{(H_{r})_{max}}$ at any $r$ quantifies the degree of
uncertainty in the knowledge of the random variable $X_{r}$. The
distribution of $f_{r}$ become completely uniform when
$\frac{H_{r}}{(H_{r})_{max}}=1$ is reached.

In the present scenario the random variables $X_{r}$ are not mutually
independent as the measurements around different centers at each
radius $r$ are carried out on the same finite volume sample and
multiple spheres around the centers can overlap. The random variables
may have extra correlations if the points are clustered or distributed
in an intrinsically inhomogeneous way. These extra correlations
increase the mutual information stored in the random variables
$X_{r}$. The decrease of information in $X_{r}$ with increasing $r$ is
evident irrespective of a homogeneous/inhomogeneous distribution but
it would diminish differently depending on the nature and the degree
of inhomogeneity present in the distribution. When inhomogeneities are
present, the relative Shannon entropy would show a departure from the
values $1$. Larger departure clearly indicates greater degree of
inhomogeneity. When the Shannon entropy $H_{r}$ attains its maximum
value $(H_{r})_{max}$ corresponding to that radius $r$ this ratio
levels up with $1$. This indicates absence of inhomogeneity beyond
that scale and may be considered as the scale of transition to
homogeneity. But this should be taken with some caution as overlap
between spheres around different centres forcibly brings down the
inhomogeneities by increasing the relative Shannon entropy at each
length scales. Overlapping become progressively more important at
larger radii. The number of available centres at a radius gradually
decreases with increasing $r$ and the centres preferentially migrate
towards the centre of the volume analyzed. The available centres are
finally confined to a smaller region when the largest spheres are
considered. This confinement bias \citep{pandey} may finally force the
relative Shannon entropy to $1$ at some radius $r$ when the spheres
completely overlap with each other. This corresponds to a situation
where the spheres around different centres show a near uniform number
count mimicking a real transition to homogeneity. The degree of
overlap as well as the confinement bias would depend on the shape of
the volume and nature of the distribution to some extent. A large
volume ensures spheres upto a larger length scales and hence may help
us to detect the scale of homogeneity if the transition occurs before
the scale where the confinement bias completely dominates the
statistics. For the present analysis one may also consider
non-overlapping spheres of different radii. But the statistics becomes
too noisy due to very small number of independent spheres at
progressively larger radii which consequently prohibits us to address
the issue of homogeneity on large scales.

The observed inhomogeneities are the combined outcome of gravitational
clustering, Poisson noise and any intrinsic inhomogeneity present in
the distribution. Inhomogeneity due to Poisson noise rapidly decreases
with $r$. The inhomogeneity due to clustering would also decrease with
$r$ albeit differently depending on the strength and nature of
clustering. Overlap causes a further suppression in the inhomogeneities
at all scales. A complete discussion on the various sources of
inhomogeneity and the factors affecting the value of the relative
Shannon entropy can be found in \citet{pandey}.

\begin{figure*}
\resizebox{14cm}{!}{\rotatebox{0}{\includegraphics{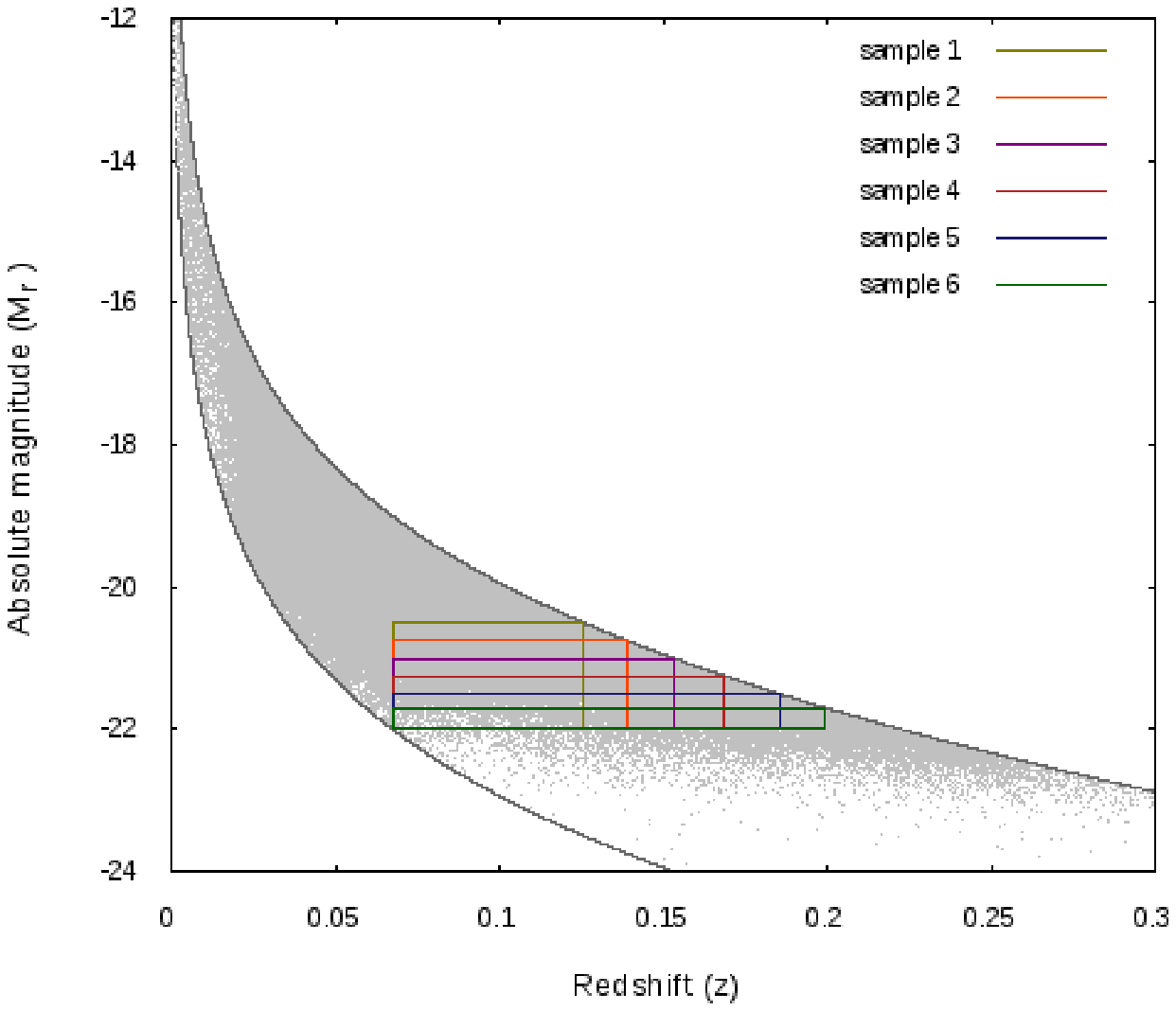}}}%
\caption{This shows the definition of our six volume limited samples
  (\autoref{tab:sdss}) from SDSS in redshift-absolute magnitude
  plane. The $6$ overlapping boxes delineate $6$ samples used in this
  analysis. The two curves define the boundary corresponding to the
  apparent magnitude limit imposed.}
  \label{fig:zm}
\end{figure*}

\begin{figure*}
 \resizebox{9cm}{!}{\rotatebox{0}{\includegraphics{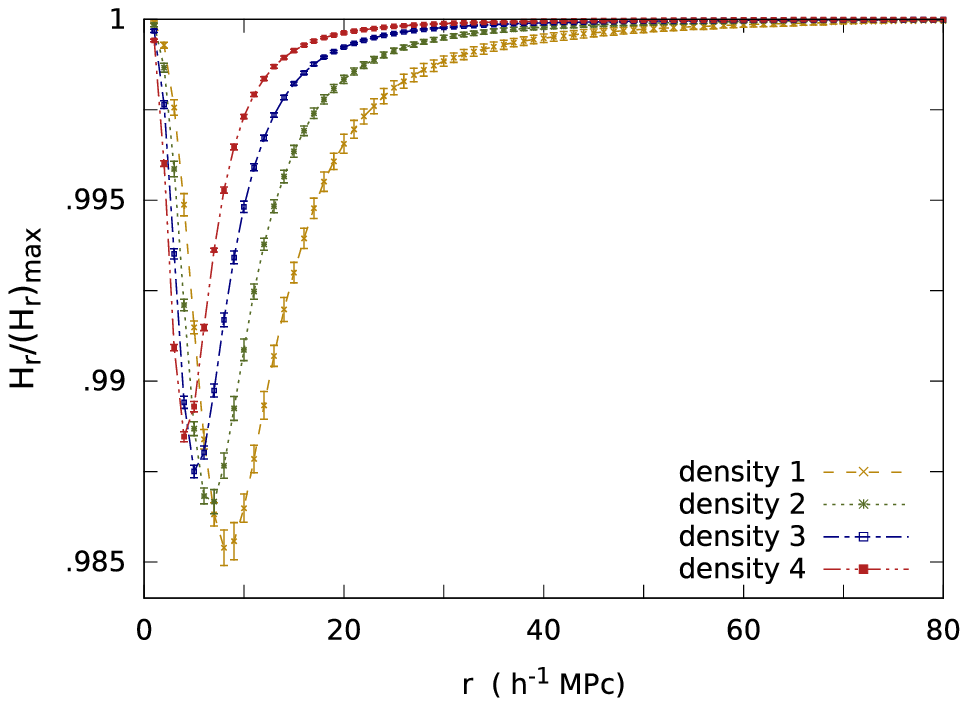}}}%
 \resizebox{9cm}{!}{\rotatebox{0}{\includegraphics{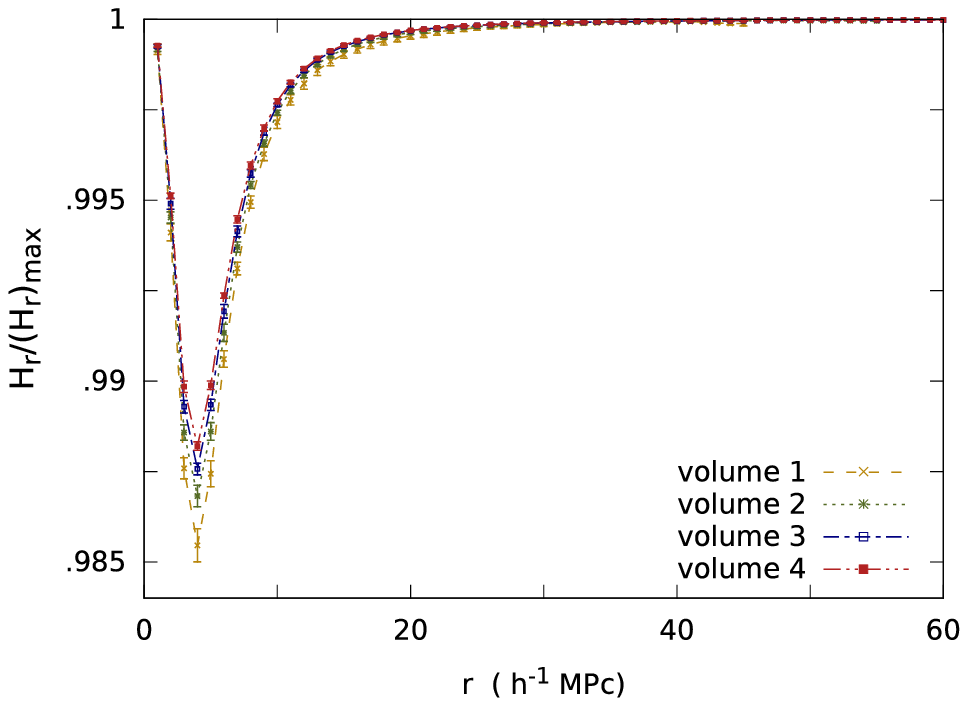}}}\\
 \resizebox{9cm}{!}{\rotatebox{0}{\includegraphics{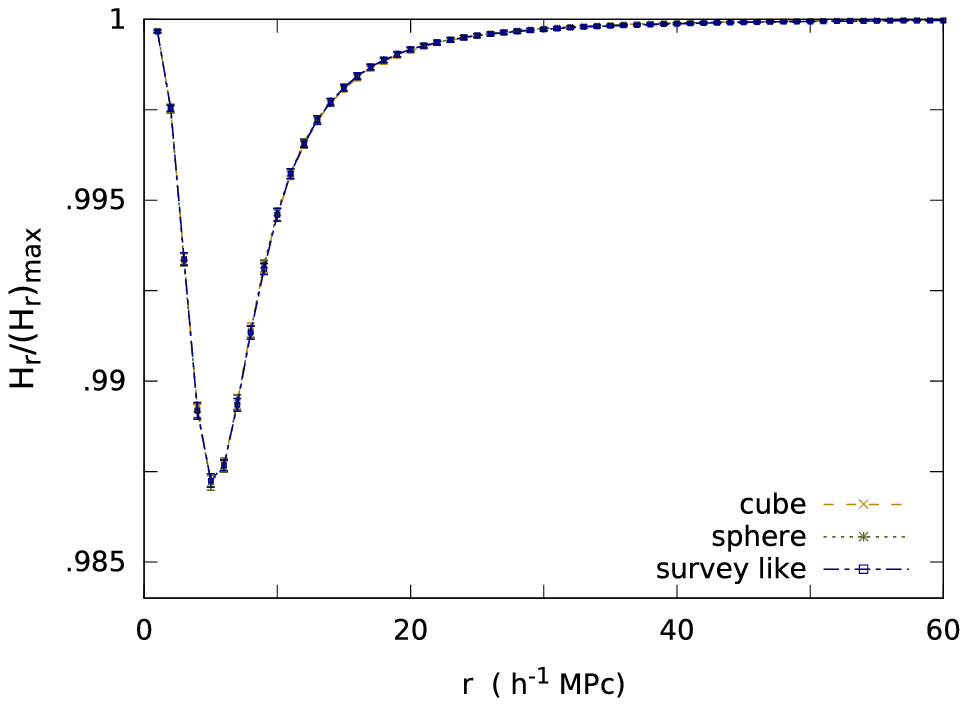}}}%
 \caption{The top left, top right and lower middle panel show the
  systematic effects of number density, volume and geometry of samples
  respectively on the relative Shannon entropy of a Poisson process at
  different length scales. The error bars shown here are the
  $1-\sigma$ variations from the $10$ realizations used in each case.}
  \label{fig:sys}
\end{figure*}

\begin{figure*}
\resizebox{9cm}{!}{\rotatebox{0}{\includegraphics{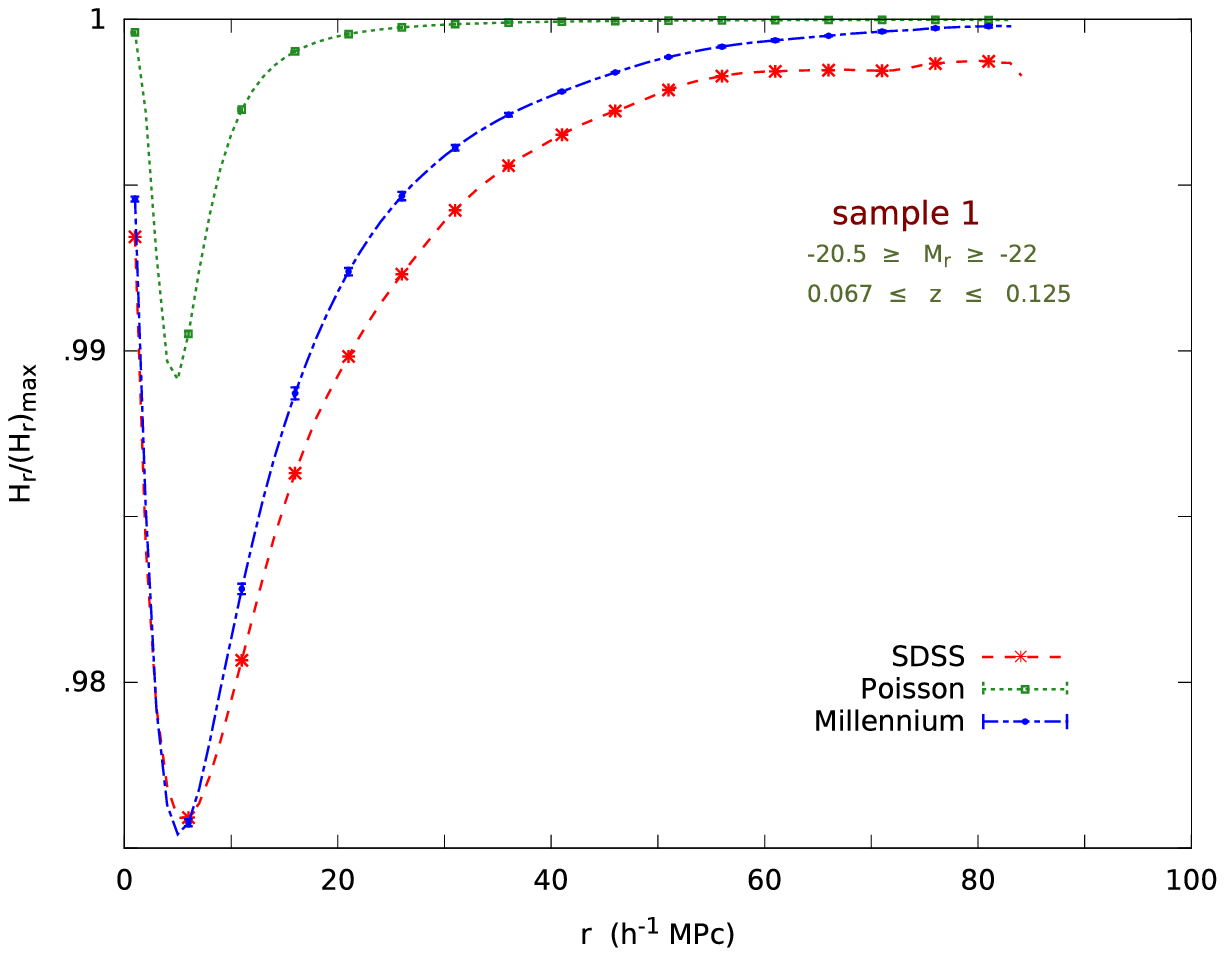}}}%
\resizebox{9cm}{!}{\rotatebox{0}{\includegraphics{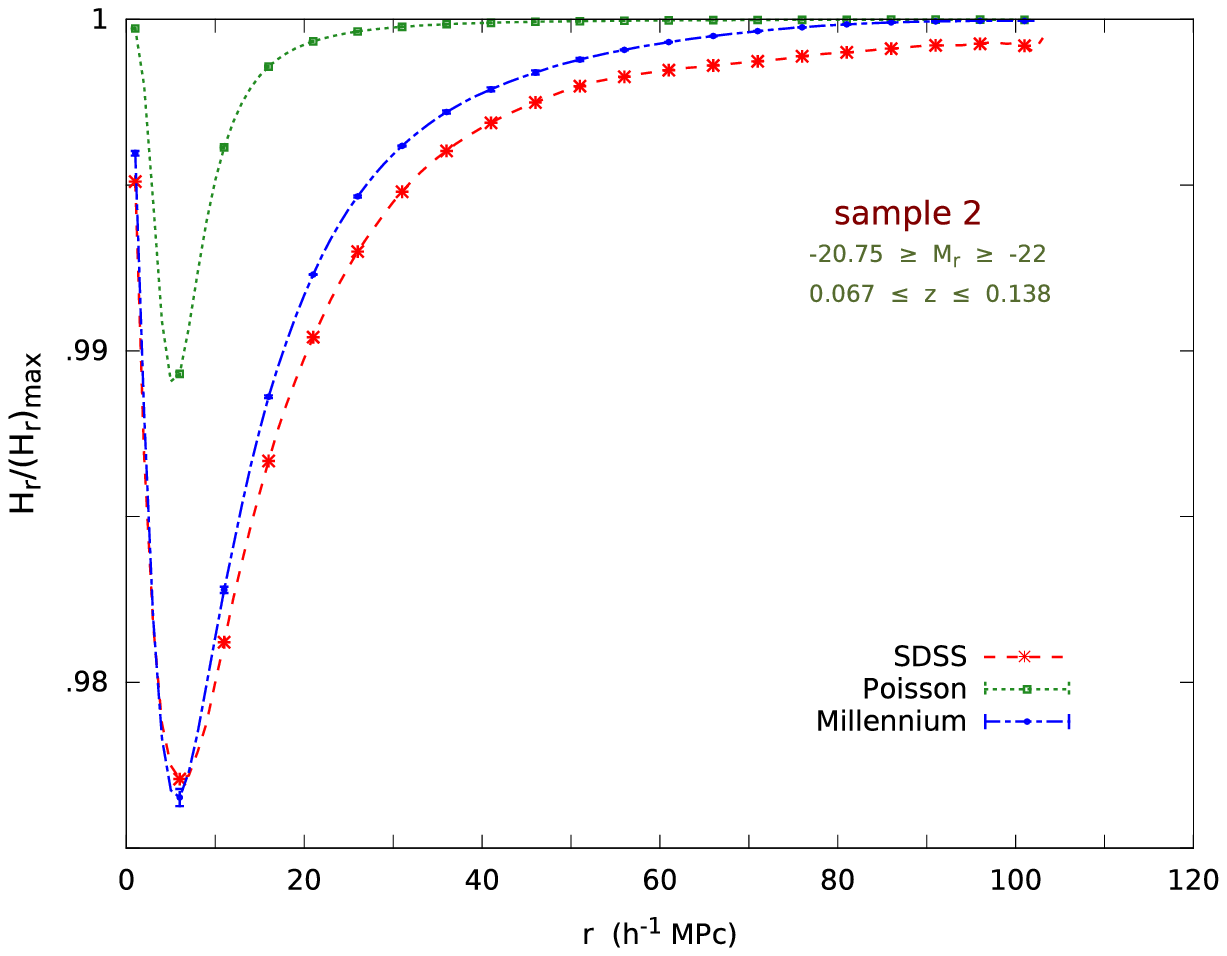}}}\\
\resizebox{9cm}{!}{\rotatebox{0}{\includegraphics{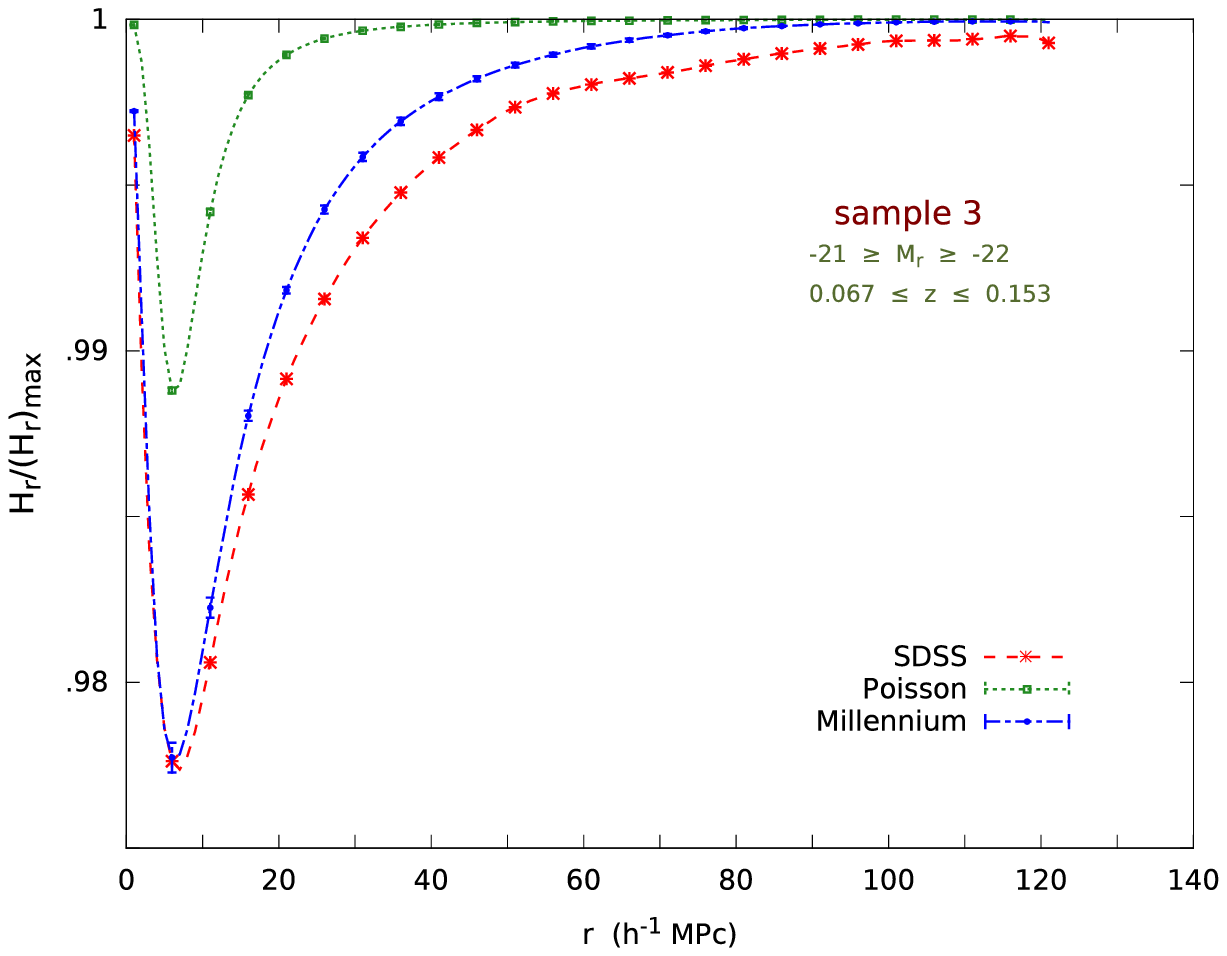}}}%
\resizebox{9cm}{!}{\rotatebox{0}{\includegraphics{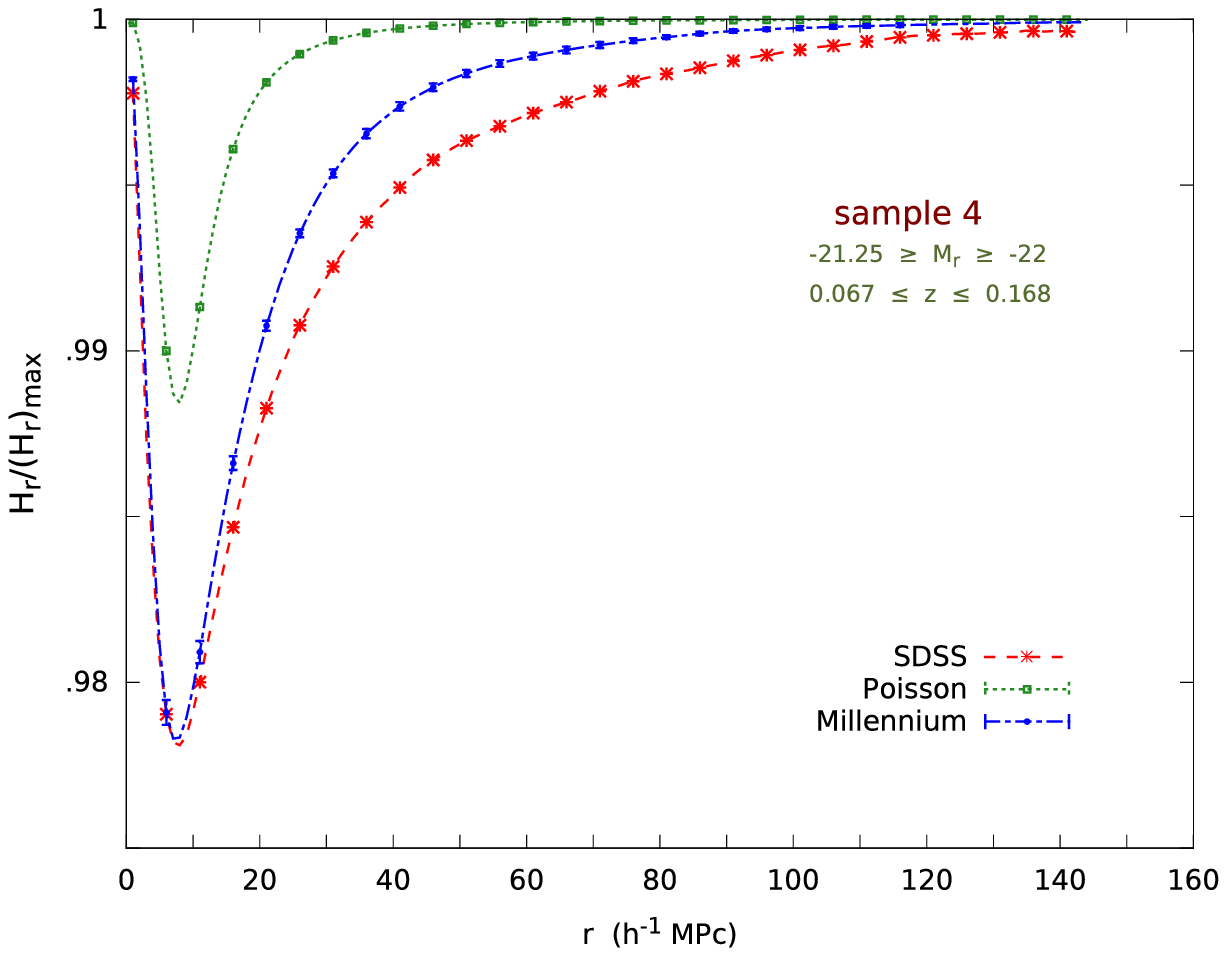}}}\\
\caption{Different panels shows the relative Shannon entropy at
  different length scales for the first four among six SDSS samples
  described in \autoref{tab:sdss}. The sample names and the
  corresponding redshift and absolute magnitude ranges are displayed
  at the top right corner of each panel. Each panel also show together
  the results for the corresponding mock galaxy samples and Poisson
  samples for comparison. The results for sample 5 and sample 6 are
  shown separately in \autoref{fig:ls} as mock samples for them could
  not be prepared from the Millennium simulation due to size
  restrictions. The error bars shown here are the $1-\sigma$
  variations from the $10$ and $3$ mock samples for the Poisson
  process and Millennium simulation respectively. The error bars are
  very tiny and nearly invisible here. The relative Shannon entropy
  for all the samples are computed at intervals of $1 \, h^{-1}\, {\rm
    Mpc}$ (as in \autoref{fig:sys}) but for clarity we show here the
  results only at intervals of $5 \,h^{-1}\, {\rm Mpc}$.}
  \label{fig:res}
\end{figure*}

\begin{figure*}
\resizebox{12cm}{!}{\rotatebox{0}{\includegraphics{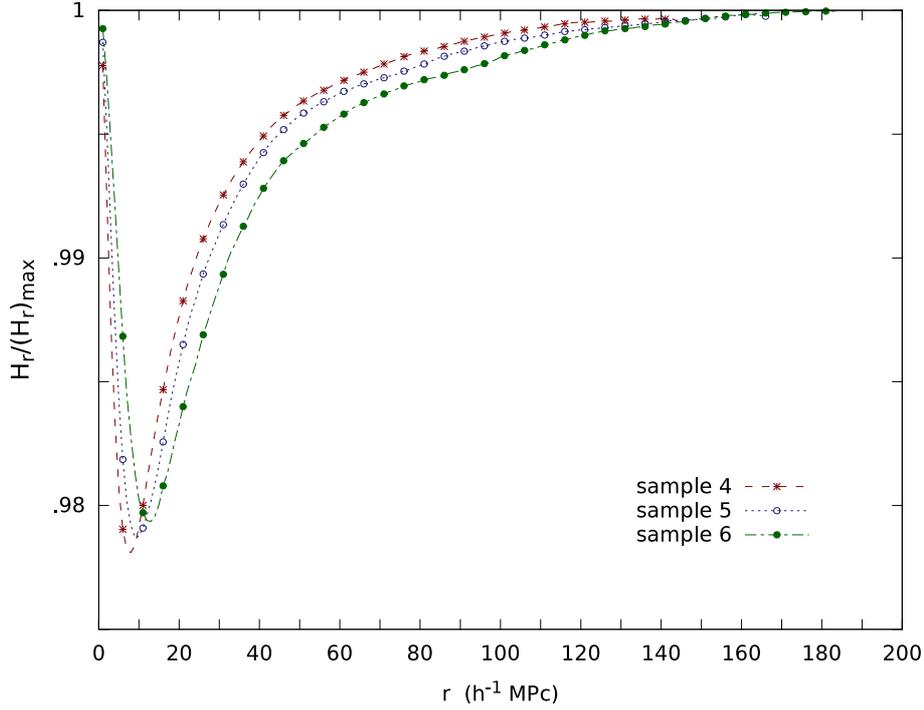}}}%
\caption{This shows the relative Shannon entropy at different length
  scales for sample 4 together with the same for two larger samples
  namely sample 5 and sample 6 described in
  \autoref{tab:sdss}. Despite having different volumes and probing
  upto different scales the results from all three samples merge
  together at $\sim 140 \,h^{-1}\, {\rm Mpc}$. The offset between the
  curves are primarily due to differences in number densities and
  clustering which cease to exit beyond $140 \,h^{-1}\, {\rm Mpc}$
  . As in \autoref{fig:res} we also show here the results at intervals
  of $5 \,h^{-1}\, {\rm Mpc}$ for clarity.}
  \label{fig:ls}
\end{figure*}

\begin{figure*}
\resizebox{12cm}{!}{\rotatebox{0}{\includegraphics{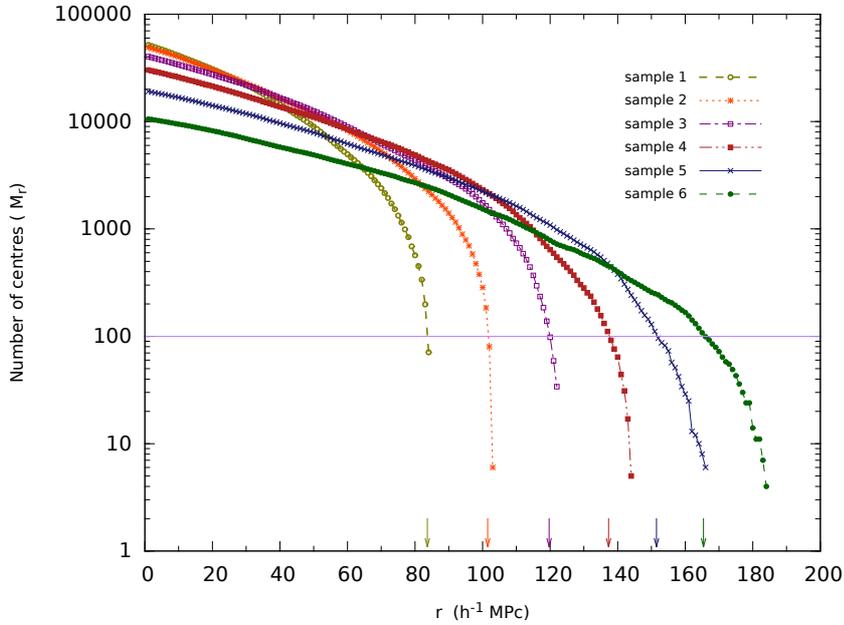}}}%
\caption{This shows the number of available centres at different
  length scales for each of the SDSS samples analyzed. The length
  scales where there are only $100$ centres for each of the SDSS
  samples are shown with arrows at the bottom. We show here the
  results at intervals of $1 \,h^{-1}\, {\rm Mpc}$.}
  \label{fig:mr}
\end{figure*}

\begin{figure*}
\resizebox{12cm}{!}{\rotatebox{0}{\includegraphics{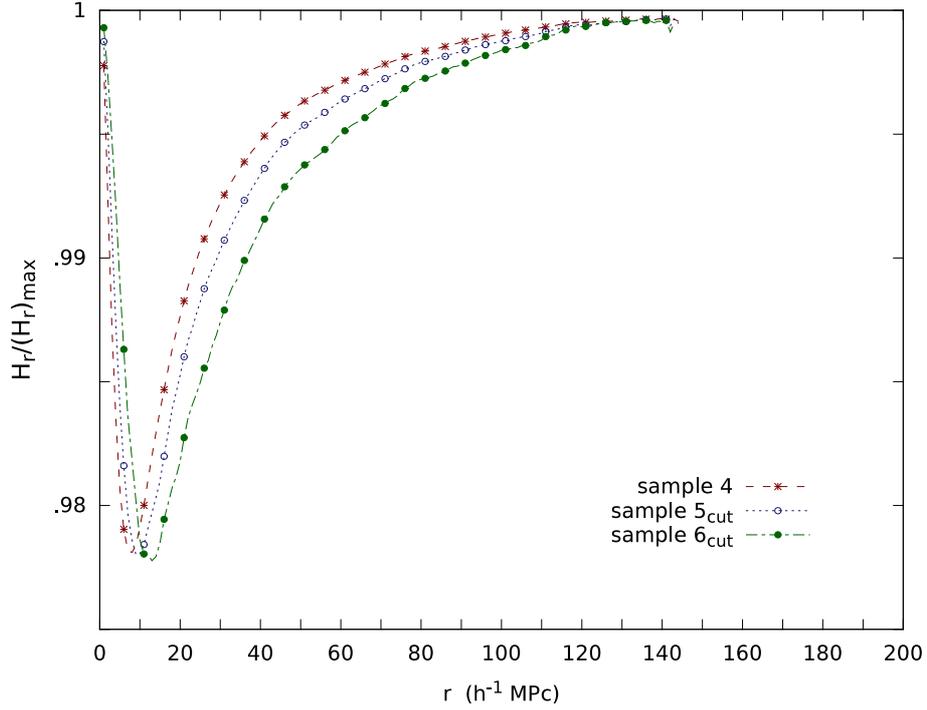}}}%
\caption{This shows the relative Shannon entropy at different length
  scales for sample 4 together with the same for sample 5 and sample 6
  after appropriate cuts are applied to them to match their sizes with
  sample 4.}
\label{fig:lsext}
\end{figure*}

\begin{figure*}
\resizebox{12cm}{!}{\rotatebox{0}{\includegraphics{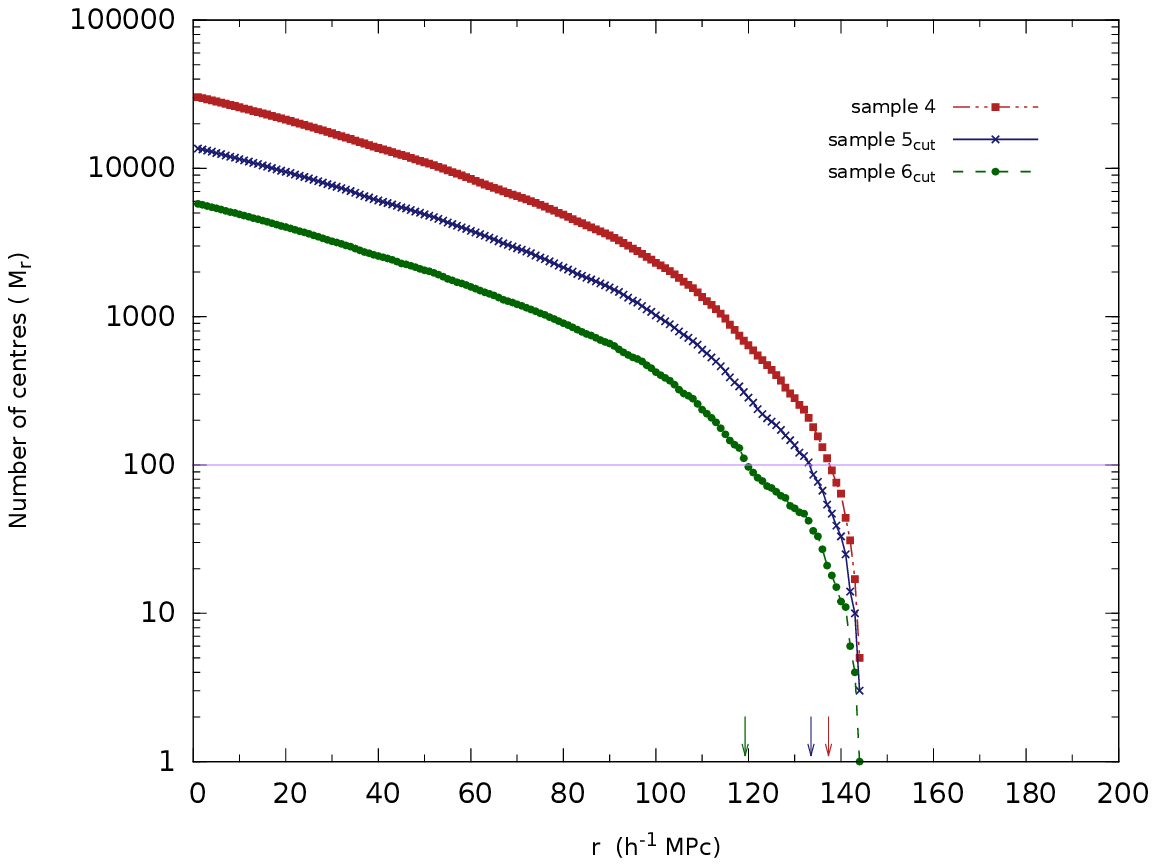}}}%
\caption{This shows the number of available centres at different
  length scales for sample 4 together with the same for sample 5 and
  sample 6 after appropriate cuts are applied to them to match their
  sizes with sample 4.}
  \label{fig:mrext}
\end{figure*}

\section{DATA}
\subsection{MONTE CARLO SIMULATIONS}

Application of the method discussed in Section 2. involves analysis of
galaxy distributions in finite regions of the Universe. Construction
of volume limited samples from galaxy surveys over different redshift
and magnitude ranges would result in samples having different volumes
and numbers of galaxies. One may also prefer to analyze volume limited
galaxy samples having different geometries. Keeping these
possibilities in mind we would like to first investigate how the
parameters like number density, volume and geometry affect the
relative Shannon entropy of a 3D spatial distribution of a point
process.

To study the possible systematic effects we generate different sets of
Monte Carlo realizations of the homogeneous Poisson point process. We
consider the following cases: (i) number density is allowed to vary
keeping the volume and geometry fixed, (ii) volume is allowed to vary
keeping the number density and geometry fixed and (iii) geometry is
allowed to vary keeping both the number density and volume fixed. We
simulate 10 Monte Carlo realizations for each density, volume and
geometry considered. \autoref{tab:den}, \autoref{tab:vol} and
\autoref{tab:geo} summarize the properties of the Monte Carlo
simulations generated to test the effects of number density, volume
and geometry respectively.

\begin{table*}{}
\caption{This shows the properties of the Monte Carlo simulations of
  homogeneous Poisson process generated inside a cube maintaining
  exactly the same volume while varying the number density. The size
  of the largest sphere tabulated here are the averages of the same
  over $10$ different realizations used in each case.}
\label{tab:den}
\begin{tabular}{|c|c|c|c|c|c|}
\hline
Sample & Box Length & Volume & No. Density & Number of & Largest Sphere\\
name & $(h^{-1}\, {\rm Mpc})$&$(h^{-1}\, {\rm Mpc})^3$ &$(h^{-1}\, {\rm Mpc})^{-3}$&points&$(h^{-1}\, {\rm Mpc})$\\
\hline
density 1&200&$ 8\times10^6$ &$5.0\times10^{-4}$&4000&95\\
density 2&200&$ 8\times10^6$ &$1.0\times10^{-3}$&8000&96\\
density 3&200&$ 8\times10^6$ &$2.0\times10^{-3}$&16000&96\\
density 4&200&$ 8\times10^6$ &$4.0\times10^{-3}$&32000&97\\
\hline
\end{tabular}
\end{table*}

\begin{table*}{}
\caption{Same as \autoref{tab:den} but here the volumes of the samples are
  allowed to change while keeping the number density and geometry
  same.}
\label{tab:vol}
\begin{tabular}{|c|c|c|c|c|c|}
\hline
Sample & Box Length & Volume & No. Density & Number of& Largest Sphere\\
name& $(h^{-1}\, {\rm Mpc})$&$(h^{-1}\, {\rm Mpc})^3$ &$(h^{-1}\, {\rm Mpc})^{-3}$&points&$(h^{-1}\, {\rm Mpc})$\\
\hline
volume 1 &100& $ 1.0\times10^6$& $5.0\times10^{-3}$ &4999 & 48\\
volume 2 &125& $ 1.95\times10^6$& $5.0\times10^{-3}$ &9765  & 60\\
volume 3 &150& $ 3.38\times10^6$& $5.0\times10^{-3}$ &16874 & 72\\
volume 4 &175& $ 5.36\times10^6$& $5.0\times10^{-3}$ &26796 & 85\\
\hline
\end{tabular}
\end{table*}

\begin{table*}{}
\caption{Same as \autoref{tab:den} but here the geometry of the samples are
  allowed to change while keeping the number density and volume
  same.}
\label{tab:geo}
\begin{tabular}{|c|c|c|c|c|c|}
\hline
Geometry & Boundary & Volume & No. Density & Number of & Largest Sphere\\
& &$(h^{-1}\, {\rm Mpc})^3$&$(h^{-1}\, {\rm Mpc})^{-3}$&points&$(h^{-1}\, {\rm Mpc})$\\
\hline
      &$0.02\leq z\leq0.08$&                   &                    &      &    \\
Survey like&$135\leq \alpha\leq225$& $ 5.82\times10^6$ & $2.0\times10^{-3}$ &11633 & 78 \\
      &$0\leq \delta\leq60$&                   &                    &      &    \\
&&&&&\\
Sphere&  radius:111.7 $h^{-1}\, {\rm Mpc}$  & $ 5.82\times10^6$ & $2.0\times10^{-3}$ &11633 & 107\\
&&&&&\\
&&&&&\\
Cube   &  side:179.8 $h^{-1}\, {\rm Mpc}$  & $ 5.82\times10^6$ & $2.0\times10^{-3}$ &11633 & 87 \\
&&&&&\\
\hline
\end{tabular}
\end{table*}

\subsection{SDSS DR12 DATA}

The Sloan Digital Sky Survey (SDSS) which started its operations with
a $2.5$m telescope \citep{gunn98, gunn06} in $2000$ has now reached
its final stage after carrying out imaging and spectroscopy over one
third of the Celestial sphere. SDSS DR12, the final data release of
SDSS \citep{alam} contains all data taken by all phases of the SDSS
through $14$th July, 2014. DR12 incorporate some significant
improvements to the spectrophotometric flux calibration. It now
contain optical spectroscopy of $2401952$ galaxies and $477161$
quasars representing a $\sim 40\%$ increase in Baryon Oscillation
Spectroscopic Survey (BOSS) spectra over the previous data release
DR10 \citep{ahn}.

Our present analysis is based on SDSS DR12 data. We have used the Main
Galaxy Sample which comprises of galaxies brighter than a limiting
r-band Petrosian magnitude $17.77$. The target selection algorithm of
the Main Galaxy Sample is detailed in \citet{strauss}. We downloaded
the data from the Catalog Archive Server (CAS) of SDSS DR12 using a
Structured Query Language (SQL) search. We identify a contiguous
region spanning $135\leq \alpha \leq 225$ and $0 \leq \delta \leq 60$
where $\alpha$ and $\delta$ are the equatorial co-ordinates. We
construct a set of volume limited samples by restricting the
extinction corrected r-band Petrosian apparent magnitude to the range
$14.5 \leq m_{r} \leq 17.5$.  The r-band absolute magnitude ranges,
the corresponding redshift ranges, number of galaxies, total volume,
number density, mean inter-particle separation and the size of the
largest sphere that can completely fit inside each of the volume
limited samples are listed in \autoref{tab:sdss}.

\begin{table*}{}
\caption{This summarizes the properties of our volume limited samples
  from SDSS.}
\label{tab:sdss}
\begin{tabular}{|c|c|c|c|c|c|c|c|c|c|}
\hline
Galaxy&Absolute& Redshift & Number of & Volume of & Number& Mean &Largest\\
Sample&magnitude range &range &Galaxies &the region  & density & separation&Sphere\\
&&&(N)&$(h^{-1}\, {\rm Mpc})^{3}$&$(h^{-1}\, {\rm Mpc})^{-3}$&$(h^{-1}\, {\rm Mpc})$&$(h^{-1}\, {\rm Mpc})$\\
\hline
&&&&&&&&&\\
sample 1&$-20.5\geq M_{r}\geq-22$&$0.067\leq z \leq0.125$&$51804$&$1.86\times10^7$&$2.77\times10^{-3}$&$7.1$&$83$\\
&&&&&&&&&\\

&&&&&&&&&\\
sample 2&$-20.75\geq M_{r}\geq-22$&$0.067\leq z \leq 0.138$&$48992$&$2.61\times10^7$&$1.87\times10^{-3}$&$8.1$&$103$\\
&&&&&&&&&\\

&&&&&&&&&\\
sample 3&$-21\geq M_{r}\geq -22$&$0.067\leq z \leq 0.153$&$40263$&$3.58\times10^7$&$1.12\times10^{-3}$&$9.6$&$122$\\
&&&&&&&&&\\

&&&&&&&&&\\
sample 4&$-21.25\geq M_{r}\geq -22$&$0.067\leq z \leq 0.168$&$30196$&$4.84\times10^7$&$6.23\times10^{-4}$&$11.7$&$144$\\
&&&&&&&&&\\

&&&&&&&&&\\
sample 5&$-21.5\geq M_{r}\geq -22$&$0.067\leq z \leq 0.185$&$19063$&$6.45\times10^7$&$2.95\times10^{-4}$&$15$&$167$\\
&&&&&&&&&\\

&&&&&&&&&\\
sample 6&$-21.7\geq M_{r}\geq -22$&$0.067\leq z \leq 0.199$&$10567$&$8.05\times10^7$&$1.31\times10^{-4}$&$19.6$&$185$\\
&&&&&&&&&\\

\hline
\end{tabular}
\end{table*}

\subsection{MILLENNIUM DATA}

Semi analytic models
\citep{white2,kauff1,kauff2,kauff3,cole1,cole2,somervil,bagh,benson,springel,guo} provide
a very powerful tool to study galaxy formation and evolution. Galaxy
formation and evolution involve many physical processes such as gas
cooling, star formation, supernovae feedback, metal enrichment,
merging and morphological evolution. The semi analytic models
parametrise the physics involved in terms of simple models following
the dark matter merger trees over time. The models provide the
statistical predictions of galaxy properties at some epoch and the
precision of these predictions are directly related to the accuracy of
the input physics. In the present analysis we use a semi-analytic
galaxy catalogue generated by \citet{guo} from the Millennium Run
simulation\citep{springel} who updated the previously available galaxy
formation models \citep{springel, croton, delucia} with improved
versions. The spectra and magnitude of the model galaxies were
computed using population synthesis models of \citet{bruzual}. We map
the galaxies to redshift space using their peculiar velocities and
then identify regions which have the same geometry as our SDSS
samples. We then apply the same magnitude cuts as those used for the
actual data and extracted the same number of galaxies as in our SDSS
samples. For each of the volume limited samples used in our analysis
we generate three mock galaxy samples from the semi analytic galaxy
catalogue.

\section{RESULTS AND CONCLUSIONS}

In the top left \autoref{fig:sys} we show the variations of the
relative Shannon entropy $\frac{H_{r}}{(H_{r})_{max}}$ with distance
$r$ for some homogeneous Poisson point processes generated inside a
cubic box with different number densities (\autoref{tab:den}). The
relative Shannon entropy versus $r$ curves show a characteristic
behaviour in all cases. At the smallest radius the relative Shannon
entropy starts from a value which is very close to $1$. It suddenly
drops at a particular length scale and then starts increasing again as
the length scale increases. The behaviour of Shannon entropy at the
smallest radius and the presence of a characteristic dip in it on a
certain length scale can be understood as follows. Any distribution
having a finite number density will look nearly uniform below the
scale of mean inter-particle separation corresponding to that
distribution. As this characteristic length scale is surpassed the
irregularities or any deviations from homogeneity can be clearly
seen. The dips in the set of curves in all panels of \autoref{fig:sys}
are located at the mean inter-particle separations of the respective
samples. While testing the effects of number density we have kept
unaltered all the other factors such as volume and the geometry of the
samples. As the density of the sample drops the characteristic dip in
the relative Shannon entropy shifts towards larger length scales (top
left panel of \autoref{fig:sys}) simply resulting from an increase in
the mean inter-particle separation. The differences in the amplitudes
of the relative Shannon entropy for Poisson samples with different
densities result from the discreteness noise or Shot noise due to
Poisson fluctuations. As the density of the sample decreases the
Poisson noise shoots up at each length scale causing a decrease in the
relative Shannon entropy and increase in the degree of inhomogeneity
at the corresponding scales. The points are uncorrelated in case of a
Poisson point process. The Void Probability function \citep{sdm} for a
homogeneous Poisson distribution is $e^{-\lambda \, V}$ where
$\lambda$ is the intensity of the Poisson process and $V$ is the
volume. It is expected that such a distribution would become
homogeneous on a scale $r \sim {\lambda}^{-\frac{1}{3}}$, which is a
measure of the average size of voids in the distribution. This is
closely related to the mean inter-particle separation and one would
expect a Poisson distribution to become homogeneous above this length
scale. But one can see in the top left panel of \autoref{fig:sys} that
the relative Shannon entropy of all the Poisson point processes
continue to show a departure from $1$ upto some distance even after
this length scale. This is due to the Poisson noise which rapidly
decreases with increasing $r$. Further the spheres around different
centres overlap with each other which reduces the inhomogeneity
increasing the relative Shannon entropy. Overlapping also becomes
progressively important at larger length scales in presence of
confinement bias. The Poisson noise and overlapping affect the
relative Shannon entropy in opposite manner in which the former
enhances whereas the later reduces inhomogeneities at each length
scales. It may be noted that the Poisson noise dominates at smaller
scales whereas overlapping dominates at larger scales. This eventually
allows the overlapping to wipe out the inhomogeneities introduced by
the Poisson noise. It can be seen in the top left panel of
\autoref{fig:sys} that the relative Shannon entropy eventually level
up with $1$ for all the Poisson point processes but at different
length scales. The lowest density sample having the largest Poisson
noise at each length scale become homogeneous on the largest length
scale among the samples analyzed. We have used here an extreme
variation in density for this test, the highest density samples being
$8$ times denser than the lowest density samples (see
\autoref{tab:den}). We also test the affects of the size of volume
occupied by the distribution. We drastically vary the volume of the
samples keeping their number density and geometry unchanged (see
\autoref{tab:vol}). The results are shown in the top right panel of
\autoref{fig:sys}. We can see that the relative Shannon entropy are
largely unaffected by the change in volume. It shows a very small
increase in relative Shannon entropy nearly on all scales, the changes
being noticeable only when the volume is increased by a factor of
$\sim 6$. This small change in Shannon entropy can be attributed to
the cosmic variance as the number of centers available at a particular
radius increases with the volume. The largest length scale that can be
probed with these samples are different as they cover different
volumes (\autoref{tab:vol}). It is interesting to note that all the
samples become homogeneous at nearly the same length scales $30 \,
h^{-1}\, {\rm Mpc}$ despite the fact that the samples with different
volumes probe upto different length scales and have different number
of centers at each radii. Although these samples cover different
volumes they have exactly the same number density ensuring the same
contribution from Poisson noise at each length scales. This clearly
demonstrates that the Poisson noise or discreteness noise governed by
the number density of the samples plays an important role in deciding
the degree of inhomogeneity and transition to homogeneity. Finally we
test the effects of geometry of the volumes on the relative Shannon
entropy. We use a survey like region, a cubic box and a spherical
region all having the same number density and volume (see
\autoref{tab:geo}). The results are shown in the lower middle panel of
\autoref{fig:sys}. We see that the degree of inhomogeneity and the
transition scale to homogeneity both are unaffected by the change in
geometry of the samples. The results of these tests clearly indicate
that the relative Shannon entropy of a distribution is most sensitive
to the number density of the distribution and nearly insensitive to
the volume and the geometry of the samples.

  In \autoref{fig:res} we show the relative Shannon entropy at
  different length scales for the first $4$ SDSS samples described in
  \autoref{tab:sdss} together with that from their mock counterparts
  from the Millennium simulation and Poisson point processes. These
  four different SDSS volume limited samples are constructed by
  keeping the brighter magnitude limit fixed at $-22$ while gradually
  shifting the fainter magnitude limit from $-20.5$ to $-21.25$ in
  steps of $-0.25$. This ensures minimal differences in the clustering
  properties of the resulting samples while allowing us to probe
  different length scales due to their different volume coverages. The
  mock samples drawn from the Millennium simulation and Poisson
  processes have identical number density, volume and geometry as
  their respective actual SDSS counterparts. We see a characteristic
  dip in the relative Shannon entropy curve for all the samples at a
  length scale which corresponds to their mean inter-particle
  separations (\autoref{tab:sdss}). Further in all the panels of
  \autoref{fig:res} we see that the galaxy samples from both the SDSS
  and the Millennium have a higher degree of inhomogeneity as compared
  to their mock Poisson samples. These differences indicate that the
  galaxy samples must have extra sources of inhomogeneity other than
  the Poisson noise resulting purely from the discrete nature of the
  distributions. It is well known that galaxy distributions exhibit
  clustering which can be quantified by their correlation
  functions. Clustering acts as an extra source of inhomogeneity for
  the galaxy samples. Besides there could be an intrinsic
  inhomogeneity built in the distribution. The inhomogeneity from
  clustering and intrinsic inhomogeneous nature if any combine
  together with that from Poisson noise to produce the total
  inhomogeneities in the distributions.  Irrespective of their nature
  inhomogeneities are expected to decrease with increasing length
  scales in presence of overlapping and confinement bias. Consequently
  these together makes it difficult to distinguish a real transition
  to homogeneity from an induced one. The transition can be confirmed
  easily if it takes place at a length scale where confinement bias
  does not dominate the statistics. For example in \autoref{fig:res}
  the mock Poisson samples corresponding to four SDSS samples show a
  transition to homogeneity in the range $30-50 \, h^{-1}\, {\rm
    Mpc}$. Although the transitions are affected by Poisson noise on
  small scales still we can make a decision on the existence of a
  genuine transition to homogeneity.  Galaxy distributions are far
  from a random Poisson distribution and if a scale of transition to
  homogeneity exist there it would possibly occur on sufficiently
  large scales where the confinement bias may overpower the
  inhomogeneities. Interestingly one can reduce the impact of
  confinement bias by simply choosing a volume which are less
  symmetric than a sphere or a cube. Although we have mentioned
  earlier that the geometry of the analyzed volume does not affect the
  relative Shannon entropy as shown in \autoref{fig:sys}, it could
  actually affect the statistics if the distribution is inhomogeneous
  nearly upto the largest length scale that can be probed with that
  volume. A volume like a rectangular slab whose two dimensions are
  significantly larger than the third one would make it sure that the
  available centres are spread across nearly the entire volume and
  would therefore never allows a situation where the spheres
  completely overlap with each other even at the largest radius. Thus
  if the number counts around different centres differ even slightly
  the relative Shannon entropy will show a deviation from $1$. The
  geometry of our SDSS samples are not exactly slab like but
  definitely has a lesser symmetry than a spherical or cubic
  volume. This may enable us to detect the signature of any
  inhomogeneities present at the largest length scales probed by these
  samples. Although overlapping would be present at each scale, it may
  not be able to erase the inhomogeneities completely even at the
  largest scale due to the lesser impact of the confinement bias. Any
  residual inhomogeneity detected at the largest length scales are
  less likely to borne out of pure Poisson noise. At the largest
  length scale the numbers of spheres could be drastically small so we
  decided to consider only upto the length scales where this number
  reduces to $100$. In \autoref{fig:mr} we show how the number of
  centers vary with length scales for all the six SDSS samples
  analyzed here. We can see that different samples probe upto
  different length scales and the scale where the number of available
  centres reduces to $100$ for sample 1, sample 2, sample 3 and sample
  4 are $83 \,h^{-1}\, {\rm Mpc}$, $102\, h^{-1}\, {\rm Mpc}$, $119 \,
  h^{-1}\, {\rm Mpc}$ and $138\, h^{-1}\, {\rm Mpc}$ respectively. In
  the top left panel of \autoref{fig:res} the relative Shannon entropy
  for the SDSS sample 1 clearly shows a deviation from unity at $80 \,
  h^{-1}\, {\rm Mpc}$ signaling some residual inhomogeneities on that
  scale. The results of the mock Poisson samples shown in
  \autoref{fig:res} indicate that for the galaxy samples having number
  densities comparable to our SDSS samples the contribution from
  Poisson noise can be safely neglected beyond $50 \, h^{-1}\, {\rm
    Mpc}$. Further sample 1 being the densest of the SDSS samples
  analyzed here would be least affected by the Poisson noise at all
  length scales. This indicates that the galaxy distribution is not
  homogeneous on $80 \, h^{-1}\, {\rm Mpc}$.  The results from SDSS
  sample 2, sample 3 and sample 4 are shown in the top right, bottom
  left and bottom right panel of \autoref{fig:res}
  respectively. sample 2, sample 3 and sample 4 are $1.4$, $2.4$ and
  $4.4$ times less denser than sample 1 (\autoref{tab:sdss}) and are
  expected to have a greater contribution from Poisson noise at all
  length scales. Interestingly the relative Shannon entropy for sample
  1, sample 2, sample 3 and sample 4 are nearly same at $80 \,
  h^{-1}\, {\rm Mpc}$. As sample 2, sample 3 and sample 4 probe
  progressively larger length scales we see that the relative Shannon
  entropy increases further with increasing length scales. It may
  appear as if the reduction in inhomogeneity with increasing length
  scales as revealed by progressively larger samples could result from
  a greater degree of overlap between the spheres. But the samples
  probing larger scales have larger volumes and for each of them we
  restrict our attention only upto a scale where the number of centers
  reduces to $100$. So the degree of overlap would not vary across the
  samples given we limit our attention upto this specific length
  scale. All the four panels of the \autoref{fig:res} show that the
  relative Shannon entropy change very slowly on large scales
  indicating that the degree of overlap do not change rapidly for our
  sample geometries possibly due to a less aggressive confinement
  bias. It may be noted further that the values of relative Shannon
  entropy for sample 2, sample 3 and sample 4 are nearly same at
  $102\, h^{-1}\, {\rm Mpc}$ and they are also same at $119 \,
  h^{-1}\, {\rm Mpc}$ for sample 3 and sample 4. These together
  clearly demonstrate the decreasing importance of Poisson noise with
  increasing length scales and indicate that any residual
  inhomogeneities at the largest length scale could be safely
  attributed to the presence of genuine inhomogeneity in the
  distribution at the corresponding length scale. We identify the
  transition scale to homogeneity corresponding to which the value of
  $1-\frac{H}{(H_{r})_{max}}$ lies within $10^{-4}$
  \citep{pandey}. This criteria is satisfied at $140 \, h^{-1}\, {\rm
    Mpc}$ for the sample 4. We also test if this result holds for
  galaxy samples having sizes bigger than sample 4 and therefore
  probing larger length scales. We construct another two volume
  limited samples namely sample 5 and sample 6 (see
  \autoref{tab:sdss}) which are larger than sample 4. We could not
  prepare the mock samples for sample 5 and sample 6 from the
  Millennium simulation due to the limitations arising out of the size
  of the simulation volume. However we analyze sample 5 and sample 6
  and find that regardless of their volumes and number densities the
  distributions become homogeneous at $\sim 140 \, h^{-1}\, {\rm
    Mpc}$. This can be clearly seen in \autoref{fig:ls} where we find
  that the value of the relative Shannon entropy of sample 4, sample 5
  and sample 6 agree within $0.02\%$ at $140 \, h^{-1}\, {\rm Mpc}$
  conforming to homogeneity. Importantly there are $\sim 400$ centers
  available at $140 \, h^{-1}\, {\rm Mpc}$ for both sample 5 and
  sample 6 which reduces to $100$ at $152 \, h^{-1}\, {\rm Mpc}$ and
  $166 \, h^{-1}\, {\rm Mpc}$ respectively. The observed offset
  between the curves below the scale of $140 \, h^{-1}\, {\rm Mpc}$
  may indicate the presence of a sample size dependent finite size
  effect. To test this we applied the same cut in the upper redshift
  limit of sample 5 and sample 6 as applied for sample 4
  (\autoref{tab:sdss}). This ensures same volume and geometry for
  these samples as they have the same lower redshift limit and sky
  coverage. We reanalyzed the resulting samples and find that the
  offset between the curves persists with very little changes in the
  differences and all three curves finally merges at $\sim 130 \,
  h^{-1}\, {\rm Mpc}$ (\autoref{fig:lsext}). The offset between the
  curves primarily originate from the differences in the number
  density and clustering in the galaxy distributions. sample 5 and
  sample 6 have brighter magnitude limits and lower number densities
  than sample 4. Poisson noise induced inhomogeneities resulting from
  different number densities plays an important role on small scales
  and become less important as the radius increases
  (\autoref{fig:sys}). Differences in inhomogeneities due to the
  differences in clustering may persist upto a large length scale. The
  fact that all three curves from sample 4, sample 5 and sample 6
  reconcile at $140 \, h^{-1}\, {\rm Mpc}$ indicates that the
  differences in their inhomogeneities cease to exist at this length
  scale. The length scale where all three curves merge reduces to
  $\sim 130 \, h^{-1}\, {\rm Mpc}$ in this case
  (\autoref{fig:lsext}). This results from the decrease in the
  available number of centres at all length scales when the sample 5
  and sample 6 are trimmed. It may be noted in \autoref{fig:mr} and
  \autoref{fig:mrext} that the number of available centers reduces
  from $\sim 1000$ to $100$ at $\sim 120 \, h^{-1}\, {\rm Mpc}$ and
  $\sim 130 \, h^{-1}\, {\rm Mpc}$ for sample 5 and sample 6
  respectively when they are trimmed to match with the size of sample
  4.

It may be interesting to note here that an analysis of filamentarity
in the Luminous Red Galaxy (LRG) distribution found that the filaments
are statistically significant upto a length scale of $120 \, h^{-1}\,
{\rm Mpc}$ \citep{pandey1}. Some recent studies \citep{maret1,maret2}
of the distribution of nearby rich galaxy clusters and superclusters
in the SDSS find the evidence of shell like structures on $120
\,h^{-1}\, {\rm Mpc}$. In different panels of \autoref{fig:res} we see
that the mock galaxy samples from the Millennium simulation also does
not become homogeneous at least up to a length scale of $80 \,
h^{-1}\, {\rm Mpc}$ and become homogeneous at $\sim 100 \, h^{-1}\,
{\rm Mpc}$. We note that all the SDSS samples exhibit a greater degree
of inhomogeneity as compared to their mock counterparts from the
Millennium simulation.  Nevertheless we can not deny that the complex
issues of overlapping and confinement bias cast some uncertainty in
the interpretations which can be resolved with yet larger samples with
a reasonable number density.

 In future we plan to carry out analysis with the Luminous Red galaxy
 distributions (LRG) \citep{eisen} and the Baryon Oscillation
 Spectroscopic Survey (BOSS) \citep{dawson} from SDSS which cover
 enormous volumes providing opportunities to probe much larger
 scales. Noticeably this would also enable us to perform the analysis
 with non-overlapping independent spheres even at sufficiently larger
 length scales. Analysis with non-overlapping spheres would have a
 remarkable advantage over the overlapping spheres as in the later the
 inhomogeneities are suppressed at each length scales making it
 difficult to detect them particularly at larger scales. This would
 help us to discern any inhomogeneities present even at the largest
 scales with nearly equal ease and settle the issue of cosmic
 homogeneity with an unprecedented confidence.

\section{ACKNOWLEDGEMENT}

Funding for SDSS-III has been provided by the Alfred P. Sloan
Foundation, the Participating Institutions, the National Science
Foundation, and the U.S. Department of Energy Office of Science. The
SDSS-III web site is http://www.sdss3.org/.

SDSS-III is managed by the Astrophysical Research Consortium for the
Participating Institutions of the SDSS-III Collaboration including the
University of Arizona, the Brazilian Participation Group, Brookhaven
National Laboratory, Carnegie Mellon University, University of
Florida, the French Participation Group, the German Participation
Group, Harvard University, the Instituto de Astrofisica de Canarias,
the Michigan State/Notre Dame/JINA Participation Group, Johns Hopkins
University, Lawrence Berkeley National Laboratory, Max Planck
Institute for Astrophysics, Max Planck Institute for Extraterrestrial
Physics, New Mexico State University, New York University, Ohio State
University, Pennsylvania State University, University of Portsmouth,
Princeton University, the Spanish Participation Group, University of
Tokyo, University of Utah, Vanderbilt University, University of
Virginia, University of Washington, and Yale University.

\bsp	
\label{lastpage}
\end{document}